\documentclass[12 pt]{amsart}
\usepackage{mathrsfs}
\usepackage{graphicx}
\usepackage{epstopdf}

 \newtheorem{thm}{Theorem}[section]
 
 \newtheorem{lem}[thm]{Lemma}
 
 \theoremstyle{definition}
 
 \theoremstyle{remark}

 \numberwithin{equation}{section}

\begin{document}
\title[Entanglement criterion for multipartite gaussian states]{Entanglement criterion independent on observables for  multipartite Gaussian states  based on uncertainty principle}

\author{Kan He}
\address[Kan He]{College of
Mathematics, Institute of Mathematics, Taiyuan University of
Technology, Taiyuan,
 030024, P. R. China} \email[K. He]{hekan@tyut.edu.cn}

\author{Jinchuan Hou}
\address[Jinchuan Hou]{College of
Mathematics, Institute of Mathematics, Taiyuan University of
Technology, Taiyuan,
 030024, P. R. China} \email[J. Hou]{houjinchuan@tyut.edu.cn}

 \maketitle

 \begin{abstract}

The local uncertainty relation (LUR) criteria for quantum
entanglement, which is dependent on chosen observables, is developed
recent. In the paper, applying the uncertainty principle, an
entanglement criteria for multipartite Gaussian states is given,
which is implemented by a minimum optimization computer program and
 independent on observalbes.
\end{abstract}

\section{Introduction}

Entanglement, as an important resource in quantum communication, has
been focused on extensively in both finite dimensional and infinite
dimensional (esp. continuous variable ) quantum systems
(\cite{Wer}-\cite{BL}). It is one of core problems to decide whether
or not a given quantum state is entangled. As we know, continuous
variable (CV) quantum systems are fundamental important from
theoretical and experimental views. In particular, Gaussian states
can be produced and managed experimentally easy, and recently the
topics on entanglement of Gaussian states have been developed
rapidly. Some different conditions for entanglement of bipartite
Gaussian states are extended from the finite dimensional case, such
as, the criterion of the positivity of the partial transpose and
additional separability criteria for covariance matrices
(\cite{Sim}, \cite{WW}), the computable cross norm (CCN) or
realignment criterion (\cite{Ru}, \cite{CHW}). The above mentioned
criteria also is generalized to the multipartite Gaussian states
(\cite{Ru2}-\cite{CW4}). Furthermore, other techniques are also used
to build the entanglement criteria for multi partite Gaussian states
(\cite{CGS}-\cite{MS}).

Entanglement criteria based on uncertainty relations has been
studied in multi partite continuous variable systems
(\cite{DGCZ}-\cite{SAWT}). Such a technique is found by Duan,
Giedke, Cirac and Zoller (\cite{DGCZ}) and the so-called local
uncertainty relation (LUR) criteria is developed by Hofmann and
Takeuchi (\cite{HT}). Roughly speaking, if one want to determine
whether or not a CV state is entangled by LUR, it needs to check
whether or not the state violates an inequality dependent on chosen
observables and parameters. For example,  Loock and Furusawa
\cite{LF} improved the LURs and says that: for an $N$-party and
$N$-mode CV state $\rho$, $\rho$ is separable if for arbitrary
scalar $h_1, h_2, ..., h_N, g_1, g_2, ..., g_N$
$$\langle(\Delta \hat{u})^2\rangle_\rho + \langle(\Delta \hat{v})^2\rangle_\rho \geq f(h_1, h_2, ..., h_N, g_1, g_2, ..., g_N),$$
where $\hat{u}=\sum_{i=1}^N h_i \hat{x}_i$, $\hat{v}=\sum_{i=1}^N
g_i \hat{p}_i$, $(\hat{x}_i, \hat{p}_i)$ is the pair of the position
and momentum operators in the $i$th mode (party) and $f$ is a
computable function. The criteria in \cite{LF} is available for
$N$-party and $N$-mode CV states (that is, there is only one mode in
each party) and dependent on observables $\hat{u}$ and $\hat{v}$. In
the present paper, we will build an operational entanglement
criterion independent on obserbables for multi-partite Gaussian
states by applying Heisenberg uncertainty principle. Applying this
criterion, it only needs to run an optimization program to determine
whether or not a Gaussian state is entangled. Compared to the
criteria in \cite{LF}, the criteria in the present paper is executed
for $N$-party  systems with arbitrary modes in each party and
independent on obserbables (See Theorem 2.1 and Corollary 2.2).

The article is organized as follows. In Section 2, we give a
sufficient condition for separability of multi partite Gaussian
states (see Theorem 2.1). Furthermore, a minimal optimization
program is designed to check entanglement of Gaussian states. In
Section 3,  kinds of pure or mixed multi-partite Gaussian states are
check by the minimal optimizer as examples. In Appendix, the proof
of Theorem 2.1 and Corollary 2.2 is given.

\section{Entanglement criteria for multi partite Gaussian states}

Let $H_1$, $H_2$, \ldots, $H_n$ be complex separable infinite
dimensional Hilbert spaces, there are $s_i$ modes in each $H_i$ for
arbitrary integer $s_i$ and $i$. Write as follows
\begin{center}
$\begin{array}{lll}
H_1:& \hat{x}_1^{(1)},\hat{x}_2^{(1)},\ldots,\hat{x}_{s_1}^{(1)}\\
& \hat{p}_1^{(1)},\hat{p}_2^{(1)},\ldots,\hat{p}_{s_1}^{(1)}
\end{array}$\\
$\begin{array}{lll}
H_2:& \hat{x}_1^{(2)},\hat{x}_2^{(2)},\ldots,\hat{x}_{s_2}^{(2)}\\
& \hat{p}_1^{(2)},\hat{p}_2^{(2)},\ldots,\hat{p}_{s_2}^{(2)}
\end{array}$\\
$\vdots$\\
$\begin{array}{lll}
H_n:& \hat{x}_1^{(n)},\hat{x}_2^{(n)},\ldots,\hat{x}_{s_n}^{(n)}\\
& \hat{p}_1^{(n)},\hat{p}_2^{(n)},\ldots,\hat{p}_{s_n}^{(n)}
\end{array}$
\end{center}
Where $(\hat{x}_j^{(i)}, \hat{p}_j^{(i)})$ is the pair of the
position and momentum operators in the $j$th mode of the $i$th
party. It is convenient to write the above notations as ${\bf\rm
common assumption }$. We have the following main result.

{\bf Theorem 2.1} {\it With the common assumption. Let
$\rho\in\mathcal S(H_1\otimes H_2 ... \otimes H_n)$ with its
covariance matrix $M_\rho=(m_{ij})_{(2 \sum s_j)\times (2 \sum
s_j)}$, $\rho$ is fully separable, then for
 two set of arbitrary real numbers $\{\alpha_j^{(i)}\}$ and
$\{\beta_j^{(i)}\}$($i=1, ..., n$ and $j=1, ..., s_i$),

$$\Gamma_{M_\rho, \alpha, \beta}=(\gamma_{k,l}(\alpha, \beta))_{n\times n}\geq 0.$$

Where
$$\begin{array}{llll} \gamma_{k,k} =& \sum_{m,h=1}^{s_k}{\alpha_m^{(k)}}{\alpha_h^{(k)}}m_{2\sum_{j=1}^{k-1}{s_j}+2m-1,2\sum_{j=1}^{k-1}{s_j}+2h-1}
 +\sum_{m,h=1}^{s_k}{\beta_m^{(k)}}{\beta_h^{(k)}}m_{2\sum_{j=1}^{k-1}{s_j}+2m,2\sum_{j=1}^{k-1}{s_j}+2h}\\ &-\sum_{i=1}^{s_k}\alpha_i^{(k)}\beta_i^{(k)}\end{array}$$
$$\gamma_{c,d}\ ( c\neq d)=\sum_{m=1}^{s_c}\sum_{h=1}^{s_d}{\alpha_m^{(c)}}{\alpha_h^{(d)}}m_{2\sum_{j=1}^{c-1}{s_j}+2m-1,2\sum_{j=1}^{d-1}{s_j}+2h-1}
 +\sum_{m=1}^{s_c}\sum_{h=1}^{s_d}{\beta_m^{(c)}}{\beta_h^{(d)}}m_{2\sum_{j=1}^{c-1}{s_j}+2m,2\sum_{j=1}^{d-1}{s_j}+2h}.$$

 }

{\bf Proof. } see the appendix.

\

The following picture is helpful to depict a sketch of
transformation from the CM $M_\rho$ to the matrix $\Gamma_{M_\rho,
\alpha, \beta}$ defined in Theorem 2.1.

$$\begin{array}{ll} \left(\begin{array}{cc|cc|cccccccccccccccccc}
     \underline{ m_{11}} &  m_{12} &  \underline{m_{13}} &  m_{14} &  \underline{m_{15}} &  m_{16} \\
       m_{21} & \underline{m_{22}} & m_{23} & \underline{m_{24}} & m_{25} & \underline{m_{26}} \\ \hline
      \underline{m_{31}} & m_{32} & \underline{m_{33}} & m_{34} & \underline{m_{35}} & m_{36} \\
      m_{41} &\underline{ m_{42}} & m_{43} & \underline{m_{44}} & m_{45} & \underline{m_{46}} \\ \hline
    \underline{ m_{51}} & m_{52} & \underline{m_{53}} & m_{54} & \underline{m_{55}} & m_{56} \\
     m_{61} &\underline{m_{62}} & m_{63} & \underline{m_{64} }& m_{65} &\underline{ m_{66}} \\
\end{array}
  \right)\end{array}_{6\times 6}\rightarrow$$
$${\small \left(
  \begin{array}{ccc}
    (\alpha_1^{(1)})^2m_{11}
 +(\beta_1^{(1)})^2m_{22}-\alpha_1^{(1)}\beta_1^{(1)} & \alpha_1^{(1)}\alpha_1^{(2)}m_{13}
 +\beta_1^{(1)}\beta_1^{(2)}m_{24} & * \\
    * & (\alpha_1^{(2)})^2m_{33}
 +(\beta_1^{(2)})^2m_{44}-\alpha_1^{(2)}\beta_1^{(2)} & * \\
    * & * & * \\
  \end{array}\right)_{3\times 3}}$$

\

Next we will design a optimization program for entanglement criteria
of Gaussian states. Firstly, we have the following corollary from
Theorem 2.1.

{\bf Corollary 2.2} {\it There exists entanglement among $H_{i_1},
H_{i_2}, ... ,H_{i_m} $ ($i_l\in \{1,2, ..., n\}, i_s\leq i_t\ {\rm
if }\ s\leq t, m\leq n$) if the scalar
$$\lambda_{i_1, i_2, ... , i_m}(M_\rho) < 0, $$ where
$$\lambda_{i_1, i_2, ... , i_m}(M_\rho)= \min_{1\leq l\leq m}
\min_{i_1\leq k\leq i_l} \min_{\{\alpha_j^{(i)}\},
\{\beta_j^{(i)}\}} |\Gamma_k(i_1, i_2, ... , i_l)|,$$
$|\Gamma_k(i_1, i_2, ... , i_l)|$ is the $k$th leading principal
minor of the  submatrix $\Gamma(i_1, i_2, ... , i_l)$ of
$\Gamma_{M_\rho, \alpha, \beta}$, $\Gamma(i_1, i_2, ... , i_l)$ is
obtained by removing the $s$th row and the $s$th column for all
$s\in \{1, 2, ..., n\}\setminus \{i_1, i_2, ... , i_l\}$. }

Applying the Corollary 2.2, we can detect entanglement of a
multi-party Gaussian state by the following optimization problem.

Let $\rho\in\mathcal S(H_1\otimes H_2 ... \otimes H_n)$ with its
covariance matrix $M_\rho=(m_{ij})_{(2 \sum s_j)\times (2 \sum
s_j)}$. To detect whether or not there exists entanglement among the
given parts $H_{i_1}, H_{i_2}, ... ,H_{i_m} $, it is the key to
minimize $|\Gamma_k(i_1, i_2, ... , i_l)|$ in Corollary 2 for fixed
$l, k$.

$$
\begin{array}{llll}
 {\rm Minimize: } & |\Gamma_k(i_1, i_2, ... , i_l)|
 \\
 {\rm Subject to: }
& \{\alpha_j^{(i)}\}\subseteq {\Bbb R}, \{\beta_j^{(i)}\}\subseteq
{\Bbb R}, (i=1, ..., n,
 j=1, ..., s_i)
\end{array} \eqno(\bf \rm OP)$$

We design the following steps to arrive at the OP problem.

S1. compute and collect leading principal minors $|\Gamma_k(i_1,
i_2, ... , i_l)|$, it is a polynomial $p(\{\alpha_j^{(i_k)}\},
\{\beta_j^{(i_k)}\})$ with $2\sum_{t=1}^k s_{i_t}$ variables
$\{\alpha_j^{(i_k)}\}, \{\beta_j^{(i_k)}\}$. The polynomial is with
constant coefficients consist of elements of $M_\rho$;

S2. compute partial derivative $\partial p/\partial
\alpha_j^{(i_k)}$ and $\partial p/\partial \beta_j^{(i_k)}$ of
$p(\{\alpha_j^{(i_k)}\}, \{\beta_j^{(i_k)}\})$ for each variable
respectively;

S3. get stationary points by solve the equation set consist of
$\partial p/\beta_j^{(i_k)}=\partial p/\beta_j^{(i_k)}=0$;

S4. compute the local minimal values of polynomial
$p(\{\alpha_j^{(i_k)}\}, \{\beta_j^{(i_k)}\})$ on all stationary
points. Finally we obtain the minimum of all local minimal values.
We solve the OP problem.

It is  mentioned that the S2 and S3 are not necessary steps to check
whether or not the Gaussian state is entangled, since the local
minimal values of polynomial $p(\{\alpha_j^{(i_k)}\},
\{\beta_j^{(i_k)}\})$ can be obtained directly by some softwares.

\section{Examples: pure or mixed states}

In this section, we check entanglement of kinds of Gaussian states,
concluding pure and mixed states.

The multi-mode pure symmetric Gaussian state is introduced in
\cite{AI}. Here we only consider the five mode case because of the
restricted space. Arbitrary a $5$-mode pure symmetric Gaussian state
has the following covariance matrix:
$$
 \left(\begin{array}{ccccccccccccccc}
a & 0& c_1 & 0 & c_1 & 0& c_1 & 0 & c_1 & 0\\
0 & a & 0 & c_2 & 0 & c_2 & 0 & c_2 & 0 & c_2\\
c_1 & 0& a & 0 & c_1 & 0 & c_1 & 0 & c_1 & 0\\
0 & c_2 & 0 & a & 0 & c_2& 0 & c_2 & 0 & c_2\\
c_1 & 0& c_1 & 0 & a & 0 & c_1 & 0 & c_1 & 0\\
0 & c_2 & 0 & c_2 & 0 & a & 0 & c_2 & 0 & c_2\\
c_1 & 0& c_1 & 0 &  c_1 & 0 &  a & 0 & c_1 & 0\\
0 & c_2 & 0 & c_2 & 0 & c_2 & 0 &  a & 0 & c_2\\
c_1 & 0& c_1 & 0 &  c_1 & 0 & c_1 & 0 & a  & 0\\
0 & c_2 & 0 & c_2 & 0 & c_2& 0 & c_2 & 0 &  a \\
\end{array}\right) \eqno(3.1)$$
Where $a\geq 1$ and
$$c_1=\frac{3(a^2-1)+\sqrt{(a^2-1)(25a^2-9)}}{8a}, c_2=\frac{3(a^2-1)-\sqrt{(a^2-1)(25a^2-9)}}{8a}.$$

We first deal with the partition $1|2|3|4|5$, that is, five modes
and five parties. In order to determine when the state $\rho_{\rm
symm}$ with the covariance matrix in Eq (3.1) is entangled, it
follows from Theorem 2.1 and Corollary 2.2 that we need to check
when the following matrix is not positive for any real scalars
$\alpha_i$ and $\beta_i$, $i=1,2,3,4,5$
$$ \left(
  \begin{array}{ccccccc}
  \begin{smallmatrix}
    a(\alpha_1^2
 +\beta_1^2)-\alpha_1\beta_1 & c_1\alpha_1\alpha_2+c_2\beta_1\beta_2 & c_1\alpha_1\alpha_3+c_2\beta_1\beta_3& c_1\alpha_1\alpha_4+c_2\beta_1\beta_4& c_1\alpha_1\alpha_5+c_2\beta_1\beta_5\\
  c_1\alpha_1\alpha_2+c_2\beta_1\beta_2 &  a(\alpha_2^2
 +\beta_2^2)-\alpha_2\beta_2 & c_1\alpha_2\alpha_3+c_2\beta_2\beta_3& c_1\alpha_2\alpha_4+c_2\beta_2\beta_4& c_1\alpha_2\alpha_5+c_2\beta_2\beta_5\\
   c_1\alpha_1\alpha_3+c_2\beta_1\beta_3 & c_1\alpha_2\alpha_3+c_2\beta_2\beta_3 &
   a(\alpha_3^2+
 +\beta_3^2)-\alpha_3\beta_3& c_1\alpha_3\alpha_4+c_2\beta_3\beta_4& c_1\alpha_3\alpha_5+c_2\beta_3\beta_5\\
  c_1\alpha_1\alpha_4+c_2\beta_1\beta_4 & c_1\alpha_2\alpha_4+c_2\beta_2\beta_4 & c_1\alpha_3\alpha_4+c_2\beta_3\beta_4& a(\alpha_4^2
 +\beta_4^2)-\alpha_4\beta_4& c_1\alpha_4\alpha_5+c_2\beta_4\beta_5\\
 c_1\alpha_1\alpha_5+c_2\beta_1\beta_5 & c_1\alpha_2\alpha_5+c_2\beta_2\beta_5 & c_1\alpha_3\alpha_5+c_2\beta_3\beta_5& c_1\alpha_4\alpha_5+c_2\beta_4\beta_5& a(\alpha_5^2 +\beta_5^2)-\alpha_5\beta_5\\
  \end{smallmatrix}
  \end{array}\right). \eqno(3.2)$$
\ \\ Set $\Gamma_i (i=1,2,3,4,5)$ are the five leading principal
minors of the matrix 3.2. Let $a=a_0$ and $a_0$ a fixed known
number, we obtain the minimal values of  $\Gamma_i$s on $\alpha_i$
and $\beta_i$, $i=1,2,3,4,5$. Only if one of $\Gamma_i$s is
negative, the Gaussian state $\rho_{\rm symm}$ is entangled under
the partition $1|2|3|4|5$. Furthermore, now if one want to ask
whether or not there exists entanglement among the mode 2, 4 and 5.
Then we only check the leading principal minors of the following
submatrix of the matrix 3.2,
$$ \left(
  \begin{array}{ccccccc}
    a(\alpha_2^2
 +\beta_2^2)-\alpha_2\beta_2 & c_1\alpha_2\alpha_4+c_2\beta_2\beta_4& c_1\alpha_2\alpha_5+c_2\beta_2\beta_5\\
  c_1\alpha_2\alpha_4+c_2\beta_2\beta_4 & a(\alpha_4^2
 +\beta_4^2)-\alpha_4\beta_4& c_1\alpha_4\alpha_5+c_2\beta_4\beta_5\\
  c_1\alpha_2\alpha_5+c_2\beta_2\beta_5 & c_1\alpha_4\alpha_5+c_2\beta_4\beta_5& a(\alpha_5^2 +\beta_5^2)-\alpha_5\beta_5\\
  \end{array}\right). \eqno(3.3)$$

Next we focus on another a partition $12|3|45$. We need to consider
the positivity of the following matrix: for any real scalar
$\alpha_i$ and $\beta_i$, $i=1,2,3,4,5$
$$ \left(
  \begin{array}{ccccccc}
  \begin{smallmatrix}
    a(\alpha_1^2+\alpha_2^2+\beta_1^2+\beta_2^2)+2c_1\alpha_1\alpha_2+2c_2\beta_1\beta_2-\alpha_1\beta_1-\alpha_2\beta_2 & c_1\alpha_3(\alpha_1+\alpha_2)+c_2\beta_3(\beta_1+\beta_2) &
    c_1(\alpha_4+\alpha_5)(\alpha_1+\alpha_2)+c_2(\beta_4+\beta_5)(\beta_1+\beta_2)\\
    c_1\alpha_3(\alpha_1+\alpha_2)+c_2\beta_3(\beta_1+\beta_2) &  a(\alpha_3^2+
 \beta_3^2)-\alpha_2\beta_2 & c_1\alpha_3(\alpha_4+\alpha_5)+c_2\beta_3(\beta_4+\beta_5)\\
    c_1(\alpha_4+\alpha_5)(\alpha_1+\alpha_2)+c_2(\beta_4+\beta_5)(\beta_1+\beta_2) & c_1\alpha_3(\alpha_4+\alpha_5)+c_2\beta_3(\beta_4+\beta_5)
     & a(\alpha_4^2+\alpha_5^2+\beta_4^2+\beta_5^2)+2c_1\alpha_4\alpha_5+2c_2\beta_4\beta_5-\alpha_4\beta_4-\alpha_5\beta_5\\
     \end{smallmatrix}
  \end{array}\right).\eqno(3.4)$$
For example, taking $a=10$, the minimal value of the determinant of
the matrix 3.4 on $\alpha_i$ and $\beta_i$ converge to $-\infty$. So
the corresponding three party Gaussian state is entangled.

For mixed Gaussian state, we consider the four modes bipartite state
with the following covariance matrix:
  $$\large
 \left(\begin{array}{cccc|ccccccccccc}
\frac{8}{5}+\lambda & \frac{2}{5}& \frac{2}{5} & \frac{2}{5} & \frac{1}{10} & \frac{1}{10} & \frac{1}{10} & \frac{1}{10}  \\
\frac{2}{5} & \frac{8}{5}+\lambda & \frac{2}{5} & \frac{2}{5} &  \frac{1}{10} & \frac{1}{10} & \frac{1}{10} & \frac{1}{10}  \\
\frac{2}{5} & \frac{2}{5}& \frac{8}{5}+\lambda & \frac{2}{5} & \frac{1}{10} & \frac{1}{10} & \frac{1}{10} & \frac{1}{10}  \\
\frac{2}{5} & \frac{2}{5} & \frac{2}{5} & \frac{8}{5}+\lambda &
\frac{1}{10} & \frac{1}{10} & \frac{1}{10} & \frac{1}{10}  \\ \hline
 \frac{1}{10} & \frac{1}{10} & \frac{1}{10} & \frac{1}{10}   & \frac{1}{2}+\lambda & -\frac{1}{8} & -\frac{1}{8} & -\frac{1}{8}\\
 \frac{1}{10} & \frac{1}{10} & \frac{1}{10} & \frac{1}{10}   & -\frac{1}{8} & \frac{1}{2}+\lambda & -\frac{1}{8} & -\frac{1}{8} \\
 \frac{1}{10} & \frac{1}{10} & \frac{1}{10} & \frac{1}{10}   &  -\frac{1}{8} & -\frac{1}{8} &  \frac{1}{2}+\lambda & -\frac{1}{8}\\
 \frac{1}{10} & \frac{1}{10} & \frac{1}{10} & \frac{1}{10}   & -\frac{1}{8} & -\frac{1}{8} & -\frac{1}{8} &  \frac{1}{2}+\lambda \\
\end{array}\right) \eqno(3.5)$$
From Theorem 2.1 and Corollary 2.2, we consider the 2$\times$2
matrix: for any real scalars $\alpha_i$ and $\beta_i$, $i=1,2,3,4$
$${\small \left(
  \begin{array}{ccccccc}
  \begin{smallmatrix}
    (\frac{8}{5}+\lambda)(\alpha_1^2+\alpha_2^2+\beta_1^2+\beta_2^2)+\frac{2}{5}\alpha_1\alpha_2+\frac{2}{5}\beta_1\beta_2-\alpha_1\beta_1-\alpha_2\beta_2 &
   \frac{1}{10}(\alpha_3+\alpha_4)(\alpha_1+\alpha_2)+\frac{1}{10}(\beta_3+\beta_4)(\beta_1+\beta_2)\\
    \frac{1}{10}(\alpha_3+\alpha_4)(\alpha_1+\alpha_2)+\frac{1}{10}(\beta_3+\beta_4)(\beta_1+\beta_2)
     & (\frac{1}{2}+\lambda)(\alpha_4^2+\alpha_5^2+\beta_4^2+\beta_5^2)-\frac{1}{8}\alpha_3\alpha_4-\frac{1}{8}\beta_3\beta_4-\alpha_4\beta_4-\alpha_5\beta_5\\
  \end{smallmatrix}
  \end{array}\right)}.\eqno(3.6)$$
For example, taking $\lambda=0.1$, the minimal value of the
determinant of the matrix (3.6) converge to $-\infty$. The bipartite
mixed Gaussian state with CM 3.5 is entangled.

\section{Conclusion}

The local uncertainty relations (LURs) is one of important classes
of entanglement criteria for the continuous variable system. It is
dependent on chosen observables. Here, we improve LURs and obtain
observable-independent entanglement criteria for arbitrary
multi-party and multi-mode Gaussian states. In particular, the
criteria can be implemented by a by a minimum optimization computer
program.

{\bf Acknowledgements} Thanks for comments. The work is supported by
National Science Foundation of China under Grant No. 11771011 and
Natural Science Foundation of Shanxi Province under Grant No.
201701D221011.

\section{Appendix }

Before the proof of Theorem 2.1, we need the following lemmas.

\begin{lem}\label{lem:1}
With the common assumption. Let \begin{equation*} \hat{X}^{(k)} =
\sum_{i=1}^{s_k}\alpha_i^{(k)}\hat{x}_i^{(k)}
\end{equation*}
\begin{equation*}
\hat{P}^{(k)} = \sum_{i=1}^{s_k}\beta_i^{(k)}\hat{p}_i^{(k)}
\end{equation*}
Then£º
\begin{equation*}
\hat{X}^{(k)}\hat{P}^{(k)}=i(\sum_{i=1}^{s_k}\alpha_i^{(k)}\beta_i^{(k)})I;
\end{equation*}
\begin{equation*}
\hat{X}^{(k)}\hat{P}^{(m)}=0,  (k\neq m);
\end{equation*}
\begin{equation*}
\hat{X}^{(k)}\hat{X}^{(m)}=\hat{X}^{(m)}\hat{X}^{(k)};
\end{equation*}
\begin{equation*}
\hat{P}^{(k)}\hat{P}^{(m)}=\hat{P}^{(m)}\hat{P}^{(k)}.
\end{equation*}
\end{lem}

\begin{lem}\label{lem:2}
With the common assumption. Let $\begin{matrix}\{t_i\}_{i=1}^n
\end{matrix}$ be a set of arbitrary real numbers. Let
\begin{equation*}
\ U = \sum_{k=1}^n t_k\hat{X}^{(k)}
\end{equation*}
\begin{equation*}
\ V = \sum_{k=1}^n t_k\hat{P}^{(k)}
\end{equation*}
 and $\rho\in S(H_1\otimes H_2\otimes ...\otimes H_n)$.
If $\rho$ is fully separable, then
\begin{equation}
\ (\Delta U)^2+(\Delta V)^2\ge
\sum_{k=1}^n(\sum_{i=1}^{s_k}\alpha_i^{(k)} \beta_i^{(k)})t_k^2
\end{equation}
\begin{proof}
$\rho$ is fully separable,
$$\rho =\int P(x)\rho_i^{(1)}\otimes\rho_i^{(2)}\otimes...\otimes\rho_i^{(n)}dx$$
\begin{equation*}
\begin{array}{lll}
\ & (\Delta U)^2+(\Delta V)^2\\ =&\int P(x)dx
[\sum_{k=1}^n(t_k^2\langle(\hat{X}^{(k)})^2\rangle_i+t_k^2\langle(\hat{P}^{(k)})^2\rangle_i)\\&
+ \sum_{l<j}2 t_l t_j\langle
\hat{X}^{(l)}\rangle_i\langle\hat{X}^{(j)}\rangle_i\\ & +
\sum_{l<j}2 t_l t_j\langle
\hat{P}^{(l)}\rangle_i\langle\hat{P}^{(j)}\rangle_i]\\ &
-\langle U\rangle_\rho^2-\langle V\rangle_\rho^2\\
= &\int P(x)dx
[\sum_{k=1}^n(t_k^2\langle(\hat{X}^{(k)})^2\rangle_i+t_k^2\langle(\hat{P}^{(k)})^2\rangle_i)\\
& +\sum_{l<j}2 t_l t_j\langle
\hat{X}^{(l)}\rangle_i\langle\hat{X}^{(j)}\rangle_i\\ & +
\sum_{l<j}2 t_l t_j\langle
\hat{P}^{(l)}\rangle_i\langle\hat{P}^{(j)}\rangle_i]\\ & -\langle
U\rangle_\rho^2-\langle V\rangle_\rho^2\\ & -\int P(x)dx
[\sum_{k=1}^n(t_k^2\langle\hat{X}^{(k)}\rangle_i^2+t_k^2\langle\hat{P}^{(k)}\rangle_i^2)]\\
&
+\int P(x)dx [\sum_{k=1}^n(t_k^2\langle\hat{X}^{(k)}\rangle_i^2+t_k^2\langle\hat{P}^{(k)}\rangle_i^2)]\\
= &\int P(x)dx
[\sum_{k=1}^n(t_k^2(\Delta\hat{X}^{(k)})_i^2+t_k^2(\Delta\hat{P}^{(k)})_i^2)]\\
& +\int P(x)dx
[(\sum_{k=1}^nt_k\langle\hat{X}^{(k)}\rangle_i)^2+(\sum_{k=1}^nt_k\langle\hat{P}^{(k)}\rangle_i)^2]\\
&
-\langle U\rangle_\rho^2-\langle V\rangle_\rho^2\\
\ge  &|\sum_{k=1}^nt_k^2\langle[\hat{X}^{(k)},\hat{P}^{(k)}]\rangle|\\
= &\sum_{k=1}^n(\sum_{k=1}^{s_k}\alpha_i^{(k)}\beta_i^{(k)})t_k^2
\end{array}
\end{equation*}
\end{proof}
\end{lem}

{\bf Proof of Theorem 2.1}

On the one hand, it follows from Lemma 5.2 that
$$
 \Delta U^2+\Delta V^2
 \geq\sum_{k-1}^n(\sum_{i=1}^{s_k}\alpha_i^k\beta_i^{k}){t_k}^2.
 $$
On the other hand,

$$\begin{array}{rl} &\Delta U^2+\Delta V^2\\= &\langle
 U^2\rangle-{\langle U\rangle}^2+\langle V^2\rangle-{\langle
V\rangle}^2
 \\&= {\langle[\sum_{i=1}^n{t_i}(\sum_{j=1}^{s_i}\alpha_j^{(i)}\hat{x}_j^{(i)})]^2\rangle}_\rho
 +{\langle[\sum_{i=1}^n{t_i}(\sum_{j=1}^{s_i}\beta_j^{(i)}\hat{p}_j^{(i)})]^2\rangle}_\rho
\\&= \sum_{i=1}^n{t_i}^2
 tr[(\sum_{m,h=1}^{s_i}\alpha_m^{(i)}\alpha_h^{(i)}\hat{x}_m^{(i)}\hat{x}_h^{(i)})\rho]
 +\sum_{i,j}^n{t_i}{t_j}
 tr[(\sum_{m=1}^{s_i}\sum_{h=1}^{s_j}\alpha_m^{(i)}\alpha_h^{(i)}\hat{x}_m^{(i)}\hat{x}_h^{(i)})\rho]
 \\&+ \sum_{i=1}^n{t_i}^2
 tr[(\sum_{m,h=1}^{s_i}\beta_m^{(i)}\beta_h^{(i)}\hat{p}_m^{(i)}\hat{p}_h^{(i)})\rho]
 +\sum_{i,j}^n{t_i}{t_j}
 tr[(\sum_{m=1}^{s_i}\sum_{h=1}^{s_j}\beta_m^{(i)}\beta_h^{(i)}\hat{p}_m^{(i)}\hat{p}_h^{(i)})\rho]
\\& =\sum_{i=1}^n{t_i}^2\sum_{m,h}^{s_i}\alpha_m^{(i)}\alpha_h^{(i)}tr(\hat{x}_m^{(i)}\hat{x}_h^{(i)}\rho)
 +\sum_{i,j}{t_i}{t_j}\sum_{m,h}^{s_i, s_j}\alpha_m^i\alpha_h^i[tr({\hat{x}_m^{(i)}}{\hat x_h^{(j)}}\rho)]
\\& +\sum_{i=1}^n{t_i}^2\sum_{m,h}^{s_i}\beta_m^{(i)}\beta_h^{(i)}tr(\hat{p}_m^{(i)}\hat{p}_h^{(i)}{\rho})
 +\sum_{i,j}{t_i}{t_j}\sum_{m,h}^{s_i, s_j}\beta_m^{(i)}\beta_h^{(j)}[tr({\hat{p}_m^{(i)}}{\hat{p}_h^{(j)}}\rho)].
\end{array}
$$
where
 $$
 \hat{x}_{m}^{i}=q_{2s_1+\cdots+2s_{i-1}+2m-1},
 \hat{p}_{m}^{i}=q_{2s_1+\cdots+2s_{i-1}+2m}
 $$
We complete the proof. \hfill$\square$

\end{document}